\documentclass[copyright,creativecommons]{eptcs}
\providecommand{\event}{LFMTP'15} 
\usepackage{color}
\usepackage{hyperref}

\definecolor{dRed}{rgb}{0.65, 0.0, 0.0}
\newcommand{\dRed}[1]{{\color{dRed}#1}}

\definecolor{dGreen}{rgb}{0.133, 0.56, 0.0}

\newcommand{\emphFact}[1]{{\dRed{#1}}}

\usepackage[sort&compress,sectionbib]{natbib}
\renewcommand{\bibname}{References}
\usepackage{todonotes}
\ifdefined\LONGVERSION
  \relax
\else
\newcommand{\LONGVERSION}[1]{}

\fi

\usepackage{listings}
\usepackage{proof}
\usepackage{multicol}
\usepackage{float}
\floatstyle{boxed}
\restylefloat{figure}
\newdimen\zzlistingsize
\newdimen\zzlistingsizedefault
\zzlistingsize=\zzlistingsizedefault
\global\def\InsideComment{0}
\newcommand{\Lstbasicstyle}{\fontsize{\zzlistingsize}{1.05\zzlistingsize}\ttfamily}
\newcommand{\keywordFmt}{\fontsize{0.9\zzlistingsize}{1.0\zzlistingsize}\bf}
\newcommand{\smartkeywordFmt}{\if0\InsideComment\keywordFmt\fi}
\newcommand{\commentFmt}{\def\InsideComment{1}\fontsize{0.95\zzlistingsize}{1.0\zzlistingsize}\rmfamily\slshape}

\newlength{\zzlstwidth}
\newcommand{\setlistingsize}[1]{\zzlistingsize=#1%
\settowidth{\zzlstwidth}{{\Lstbasicstyle~}}}
\setlistingsize{\zzlistingsizedefault}

\newtheorem{theorem}{Theorem}[section]
\newtheorem{lemma}[theorem]{Lemma}

\newtheorem{corollary}[theorem]{Corollary}

\newenvironment{proof}[1][Proof]{\begin{trivlist}
\item[\hskip \labelsep {\bfseries #1}]}{\end{trivlist}}

\newcommand{\qed}{\nobreak \ifvmode \relax \else
      \ifdim\lastskip<1.5em \hskip-\lastskip
      \hskip1.5em plus0em minus0.5em \fi \nobreak
      \vrule height0.75em width0.5em depth0.25em\fi}

\title{A Case Study on Logical Relations using Contextual Types}
\author{
Andrew Cave 
\institute{McGill University}
\institute{ Montreal QC, Canada}
\email{acave1@cs.mcgill.ca}
\and 
Brigitte Pientka
\institute{McGill University}
\institute{Montreal QC, Canada}
\email{bpientka@cs.mcgill.ca}
}

\def\titlerunning{A Case Study on Logical Relations using Contextual Types}
\def\authorrunning{A. Cave \& B. Pientka}

\begin{document}
%

\maketitle              

\begin{abstract}
Proofs by logical relations play a key role to establish rich
properties such as normalization or contextual equivalence. They are
also challenging to mechanize. In this paper, we describe the
completeness proof of algorithmic equality for simply typed
lambda-terms by Crary where we reason about logically equivalent terms
in the proof environment Beluga.  There are three key aspects we rely
upon: 1) we encode lambda-terms together with their 
operational semantics and algorithmic equality using higher-order abstract
syntax  2) we directly encode the corresponding
logical equivalence of well-typed lambda-terms using recursive types and 
higher-order functions 3) we exploit Beluga's support for contexts and
the equational theory of simultaneous substitutions. This leads to a
direct and compact mechanization, demonstrating Beluga's strength at
formalizing logical relations proofs.  
\end{abstract}

\newcommand{\R}{\mathcal{R}}
\section{Introduction}

Proofs by logical relations play a fundamental role to establish rich properties such as contextual equivalence or normalization. This proof technique goes back to Tait \cite{Tait67} and was later refined by Girard \citep{GirardLafontTaylor:proofsAndTypes}.  The central idea of logical relations is to specify relations on well-typed terms via structural induction on the syntax of types instead of directly on the syntax of terms themselves. Thus, for instance, logically related functions take logically related arguments to related results, while  logically related pairs consist of components that are related pairwise. 

Mechanizing logical relations proofs is challenging: first, specifying logical relations themselves typically requires a logic which allows arbitrary nesting of quantification and implications; second, to establish soundness of a logical relation, one must prove the Fundamental Property which says that any well-typed term under a closing simultaneous substitution is in the relation. This latter part requires some notion of simultaneous substitution together with the appropriate equational theory of composing substitutions. As Altenkirch \cite{Altenkirch:TLCA93} remarked,

\begin{quote}
``I discovered that the core part of the proof (here proving lemmas about CR) is fairly straightforward and only requires a good understanding of the paper version. However, in completing the proof I observed that in certain places I had to invest much more work than expected, e.g. proving lemmas about substitution and weakening.'' 
\end{quote}

While logical normalization proofs often are not large, they are conceptually intricate and mechanizing them has become a challenging benchmark for proof environments. There are several key questions, when we attempt to formalize such proofs: How should we represent the abstract syntax tree for lambda-terms and enforce the scope of bound variables? How should we represent well-typed terms or typing derivations? How should we deal with substitution? How can we define the logical relation on closed terms? 

Early work \cite{Berardi:WLF90,CCoquand:92,Altenkirch:TLCA93} represented lambda-terms using (well-scoped) de Bruijn indices which leads to a substantial amount of overhead to prove properties about substitutions such as substitution lemmas and composition of substitution. To improve readability and generally better support such meta-theoretic reasoning, nominal approaches support $\alpha$-renaming but substitution and properties about them are specified separately; the Isabelle Nominal package has been used in a variety of logical relations proofs from proving strong normalization for Moggi's modal lambda-calculus \citep{Doczkal:LFMTP09} to mechanically verifying the meta-theory of LF itself including the completeness of equivalence checking \citep{Narboux:LFMTP08,Urban:TOCL11}. 

Approaches representing lambda-terms using higher-order abstract syntax (HOAS) trees (also called $\lambda$-tree syntax) model binders in the object language (i.e. in our case the simply typed lambda-calculus) as binders in the meta language (i.e. in our case the logical framework LF \citep{Harper93jacm}). Such encodings inherit not only $\alpha$-renaming and substitution from the meta-language, but also weakening and substitution lemmas. However, direct encodings of logical relations proofs is beyond the logical strength supported in systems such as Twelf \citep{Pfenning99cade}. In this paper, we demonstrate the power and elegance of logical relations proofs within the proof environment Beluga \citep{Pientka:IJCAR10} which is built on top of the logical framework LF. Beluga allows programmers to pair LF objects together with their surrounding context and 
this notion is internalized  as a contextual type $[\Psi \vdash A]$ which is inhabited by term $M$ of type $A$ in the context $\Psi$ \citep{Nanevski:ICML05}.  Proofs about contexts and contextual LF objects are then implemented as dependently-typed recursive functions via pattern matching \citep{Pientka:POPL08,Pientka:PPDP08}.  Beluga's functional language supports higher-order functions and indexed recursive data-types \citep{Cave:POPL12} which we use to encode the logical relation. As such it does not impose any restrictions as for example found in Twelf \citep{Pfenning99cade} which does not support arbitrary quantifier alternation or Delphin \citep{Schuermann:ESOP08} which lacks recursive data-types. Recently, Beluga has been extended to first-class simultaneous substitutions allowing abstraction over substitutions and supporting a rich equational theory about them \citep{Cave:LFMTP13,Pientka:CADE15}.  

In this paper, we describe the completeness proof of algorithmic equality for simply typed
lambda-terms by Crary \cite{Crary:ATAPL} where we reason about logically equivalent terms
in the proof environment Beluga.  There are three key aspects we rely
upon: 1) we encode lambda-terms together with their 
operational semantics together with algorithmic equality using higher-order abstract syntax 
2) we directly encode the corresponding logical equivalence of well-typed lambda-terms using recursive  types and higher-order functions 3) we exploit Beluga's support for contexts and
the equational theory of simultaneous substitutions. This leads to a
direct and compact mechanization and allows us to demonstrate Beluga's strength at
formalizing logical relations proofs. Based on this case study we also draw some general lessons.

\section{Proof Overview: Completeness of Algorithmic Equality}\label{sec:proofoverview}
In this
section we give a brief overview of the motivation and high level structure of
the completeness proof of algorithmic equality. For more detail, we refer the
reader to \cite{Crary:ATAPL} and \cite{Harper03tocl}. Extensions of 
this proof are important for the metatheory of dependently typed systems such as
LF and varieties of Martin-L\"of 
Type Theory, where they are used to establish decidability of typechecking. The proof concerns
three judgements, the first of which is declarative equivalence:

\begin{center}
  \begin{tabular}{@{}l@{~~~}l@{}}
$\Gamma \vdash M \equiv N : A$ & terms M and N are declaratively equivalent at
type $A$
  \end{tabular}
\end{center}

Declarative equivalence includes convenient but non-syntax directed rules
such as transitivity and symmetry, among rules for congruence,
extensionality and $\beta$-contraction. We will see the full definition
in Sec. \ref{sec:mechanization}. In particular, it may include
apparently type-directed rules such as extensionality at unit type:

$$
\infer{\Gamma \vdash M \equiv N : \mathrm{Unit}}{\Gamma \vdash M : \mathrm{Unit} &
  \Gamma \vdash N : \mathrm{Unit}}
$$

This rule relies crucially on type information, so the common
untyped rewriting strategy for deciding equivalence no longer
applies. Instead, one can define an algorithmic notion of equivalence
which is directed by the syntax of types. This is the path we follow
here. We define algorithmic term equivalence 
mutually with path equivalence, which is the syntactic equivalence of terms headed by
variables, i.e. terms of the form $x\,M_1\,...\,M_n$. 

\begin{center}
  \begin{tabular}{@{}l@{~~~}l@{}}
$\Gamma \vdash M \Leftrightarrow N : A$ & terms M and N are
algorithmically equivalent at type $A$ \\
$\Gamma \vdash M \leftrightarrow N : A$ & paths M and N are
algorithmically equivalent at type $A$ \\
  \end{tabular}
\end{center}

In what follows, we sketch the proof of completeness of
algorithmic equivalence for declarative 
equivalence. A direct proof by induction over
derivations fails unfortunately in the application case where we need to show that
applying equivalent terms to equivalent arguments yields equivalent
results, which is not so easy. Instead, one can proceed by proving a
more general statement that declaratively equivalent terms are
\emph{logically equivalent}, and so in turn algorithmically
equivalent. Logical equivalence is a relation defined directly on the
structure of the types. We write it as follows:

\begin{center}
  \begin{tabular}{@{}l@{~~~}l@{}}
$\Gamma \vdash M \approx N : A$ & Terms M and N are logically equivalent at
type $A$ \\
  \end{tabular}
\end{center}

The key case is at function type, which directly defines logically
equivalent terms at function type as taking logically equivalent arguments
to logically equivalent results. Crary defines:

\begin{center}
  \begin{tabular}{@{}l@{~~~}l@{~~~}l@{}}
$\Gamma \vdash M_1 \approx M_2 : A \Rightarrow B$ & iff & \emphFact{for all $\Delta \geq \Gamma$}
and $N_1$, $N_2$, \\
& & if $\Delta \vdash N_1 \approx N_2 : A$\\
& & then $\Delta \vdash M_1\; N_1 \approx M_2\; N_2 : B$
  \end{tabular}
\end{center}

A key complication is the quantification over all extensions $\Delta$
of the context $\Gamma$. This is essential to show completeness of the
algorithmic rule for function types, which states that that to compare
two terms $\Gamma \vdash M \Leftrightarrow N : A \Rightarrow B$ it suffices to
compare their applications to \emph{fresh} variables: $\Gamma,x:A
\vdash M x \Leftrightarrow N x : B$. The
generalization to \emph{all} extensions $\Delta$ of $\Gamma$ then
arises naturally. This Kripke-style monotonicity condition
is one of the reasons that this proof is more challenging than normalization
proofs for simply typed lambda-terms, where this
quantification can often be avoided using other technical tricks.

For our formalization, we take a slightly different approach which 
better exploits the features of Beluga available to us. We
instead quantify over an arbitrary context $\Delta$ together with a
simultaneous substitution $\pi$ which provides for each $x{:}T$ in
$\Gamma$, a path $M$ satisfying $\Delta \vdash M \leftrightarrow M :
T$. We will call such a substitution a \emph{path substitution} and write this
condition as $\Delta \vdash \pi : \Gamma$. In the course of the completeness proof, $\pi$
will actually only ever be instantiated by substitutions which simply
perform weakening. That is, $\Delta$ will be of the form
$\Gamma,\Gamma'$ where $\Gamma = x_1{:}A_1,...,x_n{:}A_n$ and $\pi$ will be of the form
$\Gamma,\Gamma' \vdash x_1/x_1,...,x_n/x_n : \Gamma$. However, the extra generality of path
substitutions surprisingly does no harm to the proof, and fits well within Beluga.

\begin{center}
  \begin{tabular}{@{}l@{~~~}l@{~~~}l@{}}
$\Gamma \vdash M_1 \approx M_2 : A \Rightarrow B$ & iff & \emphFact{for all
$\Delta$, path substitutions $\Delta
\vdash \pi : \Gamma$}, and $N_1$, $N_2$ \\
& & if $\Delta \vdash N_1 \approx N_2 : A$ \\
& & then \emphFact{$\Delta \vdash M_1[\pi]\; N_1 \approx M_2[\pi]\; N_2 : B$}
  \end{tabular}
\end{center}

The high level goal is to establish that declaratively equivalent
terms are logically equivalent, and that logically equivalent terms
are algorithmically equivalent. The proof requires establishing a
few key properties of logical equivalence. The first is
monotonicity, which is crucially used for weakening logical
equivalence. This is used when applying terms to fresh variables.

\begin{lemma}[Monotonicity]$\;$\\
If $\Gamma \vdash M \approx N : A$ and $\Delta \vdash \pi : \Gamma$ is
a path substitution, then $\Delta \vdash M[\pi] \approx N[\pi] : A$
\end{lemma}

The second key property is (backward) closure of logical equivalence
under weak head reduction. This is proved by induction on the type $A$.

\begin{lemma}[Logical weak head closure]$\;$\\
If $\Gamma \vdash N_1 \approx N_2 : A$  and $M_1 \longrightarrow^*_{wh} N_1$
and $M_2 \longrightarrow^*_{wh} N_2$ then $\Gamma \vdash M_1 \approx M_2 : A$
\end{lemma}

In order to escape logical equivalence to obtain algorithmic
equivalence in the end, we need the main lemma, which is a mutually
inductive proof showing that path equivalence is included in
logical equivalence, and logical equivalence is included in
algorithmic equivalence:

\begin{lemma}[Main lemma]$\;$
\begin{enumerate}
\item If $\Gamma \vdash M \leftrightarrow N : A$ then $\Gamma \vdash M
  \approx N : A$
\item If $\Gamma \vdash M \approx N : A$ then $\Gamma \vdash M
  \Leftrightarrow N : A$
\end{enumerate}
\end{lemma}

Also required are symmetry and transitivity of logical equivalence,
which in turn require symmetry and transitivity of algorithmic
equivalence, determinacy of weak head reduction, and uniqueness of
types for path equivalence. We will not go
into detail about these lemmas, as they are relatively mundane, but refer the
reader to the discussion in \cite{Crary:ATAPL}.

What remains is to show that declarative equivalence implies logical
equivalence. This requires a standard technique to generalize the
statement to all instantiations of open terms by related substitutions. If
$\sigma_1$ is of the form $M_1/x_1,...,M_n/x_n$ and $\sigma_2$ is of
the form $N_1/x_1,...,N_n/x_n$ and $\Gamma$ is of the form
$x_1{:}A_1,...,x_n{:}A_n$, we write
$\Delta \vdash \sigma_1 \approx \sigma_2 : \Gamma$ to mean that $\Delta \vdash M_i \approx N_i : A_i$ for all $i$.

\begin{theorem}[Fundamental theorem]$\;$\\
If $\Gamma \vdash M \equiv N : A$  and $\Delta \vdash \sigma_1 \approx
\sigma_2 : \Gamma$ then $\Delta \vdash M[\sigma_1] \approx
N[\sigma_2] : A$
\end{theorem}

The proof goes by induction on the derivation of $\Gamma \vdash M
\equiv N : A$. We show one interesting case in order to demonstrate some sources
of complexity.

\begin{proof}
Case: $\infer{\Gamma \vdash \lambda x. M_1 \equiv \lambda x. M_2 : A
  \Rightarrow B}{\Gamma,x:A \vdash M_1 \equiv M_2 : B}$
\begin{enumerate}
\item Suppose we are given $\Delta'$, a path substitution $\Delta' \vdash \pi : \Delta$ and $N_1,N_2$ with $\Delta' \vdash N_1 \approx N_2 :
A$.
\item We have $\Delta' \vdash \sigma_1[\pi] \approx \sigma_2[\pi] : \Gamma$
\hfill (by monotonicity)
\item Hence $\Delta' \vdash (\sigma_1[\pi],N_1/x) \approx
  (\sigma_2[\pi],N_2/x) : \Gamma,x:A$ \hfill  (by definition)
\item Hence $\Delta' \vdash M_1[\sigma_1[\pi],N_1/x] \approx
  M_2[\sigma_2[\pi],N_2/x] : B$ \hfill  (by induction hypothesis)
\item Hence $\Delta' \vdash M_1[\sigma_1[\pi],x/x][N_1/x] \approx
  M_2[\sigma_2[\pi],x/x][N_2/x] : B$ \hfill  (by substitution properties)
\item Hence $\Delta' \vdash (\lambda x. M_1[\sigma_1[\pi],x/x])\; N_1
  \approx (\lambda x. M_2[\sigma_2[\pi],x/x])\; N_2 : B$ \hfill  (by weak head
  closure)
\item Hence $\Delta' \vdash ((\lambda x. M_1)[\sigma_1])[\pi]\; N_1
  \approx ((\lambda x. M_2)[\sigma_2])[\pi]\; N_2 : B$ \hfill  (by substitution properties)
\item Hence $\Delta \vdash (\lambda x. M_1)[\sigma_1] \approx (\lambda
  x. M_2)[\sigma_2] : A \Rightarrow B$ \hfill  (by definition of logical equivalence)
\end{enumerate}
\end{proof}

We observe that this proof relies heavily on equational properties of
substitutions. Some of this complexity appears to be due to our choice of
quantifying over substitutions $\Delta \vdash \pi : \Gamma$ instead of
extensions $\Delta \geq \Gamma$. However,
we would argue that reasoning instead about extensions $\Delta \geq
\Gamma$ does not remove this complexity, but only rephrases it.

Finally, by establishing the relatedness of the identity substitution to
itself, i.e. $\Gamma \vdash \mathrm{id} \approx \mathrm{id} : \Gamma$ we can combine the
fundamental theorem with the main lemma to obtain completeness.

\begin{corollary}[Completeness]
If $\Gamma \vdash M \equiv N : A$ then $\Gamma \vdash M
\Leftrightarrow N : A$
\end{corollary}

\section{Mechanization}\label{sec:mechanization}
We mechanize the development of the declarative and algorithmic equivalence
together with its completeness proof in Beluga, a dependently typed proof
language built on top of the logical framework LF. The central idea is to
specify lambda-terms, small-step semantics, and type-directed
algorithmic equivalence in the logical framework LF. This allows us to model bindings uniformly using the LF
function space and obviates the need to model and manage names
explicitly. Beluga's proof language allows programmers to encapsulate LF objects
together with their surrounding context as contextual objects and provides
support for higher-order functions, indexed recursive types, and pattern
matching on contexts and contextual objects. We define logical
equivalence and (for technical reasons) declarative equivalence using
indexed recursive types. All our proofs will then be implemented as recursive
functions using pattern matching and pass the totality checker. The complete source code for our
development can be found in the directory \lstinline!examples/logrel! of the Beluga
distribution which is available at 
\url{https://github.com/Beluga-lang/Beluga}.

\subsection{Encoding lambda-terms, typing and reduction in the logical framework
LF}
Our proof is about a simply-typed lambda calculus with one base type
\lstinline{i}. Extending the proof to support a unit type and products
is straightforward. We describe the types and terms in LF as follows, employing
HOAS for the representation of lambda abstraction. That is, we express
the body of the lambda expression as an LF function
\lstinline{tm -> tm}. There is no explicit case for variables; they
are implicitly handled by LF. We show side by side the corresponding grammar.

\begin{minipage}{7cm}
\begin{lstlisting}
LF tp : type =
| i :  tp
| $\Rightarrow$ : tp -> tp -> tp % infix
;
LF tm : type =
| app : tm -> tm -> tm
| lam : (tm -> tm) -> tm;
\end{lstlisting}  
\end{minipage}
\begin{minipage}{6cm}
\[ \begin{array}{l@{~~}lll}
\mbox{Types} & T,S & ::= & i \mid T \Rightarrow S \\ \\[1em]
\mbox{Terms} & M,N & ::= & x \mid \mathrm{lam}\, x.\, M \mid \mathrm{app}\, M \,N\\[1em]\;
\end{array} \]  
\end{minipage}

Finally, we describe also weak head reduction for our terms. Notice here that
the substitution of \lstinline{N} into \lstinline{M} in the
$\beta$-reduction case is accomplished using LF application. We then
describe multi-step reductions as a sequence of single step
reductions. All free variables occurring in the LF signature are reconstructed
and bound implicitly at the outside.

\begin{lstlisting}
LF step : tm -> tm -> type =
| beta    : step (app (lam M) N) (M N)
| stepapp : step M M' -> step (app M N) (app M' N);

LF mstep : tm -> tm -> type =
| refl   : mstep M M
| trans1 : step M M' -> mstep M' M'' -> mstep M M'';
\end{lstlisting}

\subsection{Encoding algorithmic equivalence}
We now describe the algorithmic equality of terms. This is defined
as two mutually recursive LF specifications. We write
\lstinline{algeqTm M N T}
for algorithmic equivalence of terms \lstinline{M} and \lstinline{N} at
type \lstinline{T} and 
\lstinline{algeqP P Q T} for algorithmic path equivalence at type
\lstinline{T} -- these are terms whose head is a variable, not a lambda abstraction. Term equality is directed by the type,
while path equality is directed by the syntax. Two terms \lstinline{M}
and \lstinline{N} at base type \lstinline{i} are equivalent if they weak head reduce to weak head normal
terms \lstinline{P} and \lstinline{Q} which are path equivalent. Two
terms \lstinline{M} and \lstinline{N} are equivalent at type
\lstinline{T $\Rightarrow$ S} if applying them to a fresh variable
\lstinline{x} of type \lstinline{T} yields equivalent terms. Variables are only path equivalent to
themselves, and applications are path equivalent if the terms at
function position are path equivalent, and the terms at argument
positions are term equivalent. 

\begin{lstlisting}
LF algeqTm: tm -> tm -> tp -> type = 
| algbase: mstep M P -> mstep N Q -> algeqP P Q i -> algeqTm M N i.
| algarr : ({x:tm} algeqP x x T -> algeqTm (app M x) (app N x) S) -> algeqTm M N (arr T S)
and algeqP : tm -> tm -> tp -> type
| algapp : algeqP M1 M2 (arr T S) -> algeqTm N1 N2 T -> algeqP (app M1 N1) (app M2 N2) S;
\end{lstlisting}

By describing algorithmic equality in LF, we gain structural
properties and substitution for free. For this particular proof, only weakening
is important.

A handful of different forms of contexts are relevant for this proof. We describe these
with \lstinline{schema} definitions in Beluga. Schemas classify contexts in a
similar way as LF types classify LF objects. Although schemas are
similar to Twelf's world declarations, schema checking does 
not involve verifying that a given LF type family only introduces the
assumptions specified in the schema; instead schemas will be used by the
computation language to guarantee that we are manipulating contexts of a certain
shape. Below, we define the schema \lstinline{actx}, which enforces that term
variables come paired with an algorithmic equality assumption
\lstinline{algeqP x x t} for some type \lstinline{t}.

\begin{lstlisting}
schema actx = some [t:tp] block x:tm, ax:algeqP x x t;
\end{lstlisting}

\subsection{Encoding logical equivalence}
To define logical equivalence, we need the notion of 
path substitution mentioned in
Sec. \ref{sec:proofoverview}. For this purpose, we use Beluga's
built-in notion of simultaneous substitutions. We write \lstinline{[$\delta \vdash \gamma$]} for the
built-in type of simultaneous substitutions which provide for each
variable in the context $\gamma$ a corresponding term in the context
$\delta$. When $\gamma$ is of schema \lstinline{actx}, such a
substitution consists of blocks of the form \lstinline{M/x,P/ax} where
\lstinline{M} is a term and \lstinline{P} is a derivation of \lstinline{algeqP M M T}, just as we need.

To achieve nice notation, we define an LF type of
pairs of terms, where the infix operator $\approx$ simply constructs a pair
of terms:

\begin{lstlisting}
LF tmpair : type = 
| $\approx$ : tm -> tm -> tmpair % infix;
\end{lstlisting}

Logical equivalence, written \lstinline{Log [$\gamma$ |- M $\approx$ N] [A]},
expresses that \lstinline!M! and \lstinline!N! are logically related
in context $\gamma$ at type $A$. We embed contextual objects into
computations and recursive types wrapping them inside
 \lstinline![ ]!. Since \lstinline!M! and \lstinline!N! are used in
 the context $\gamma$, by default, they can depend on
 $\gamma$. Formally, each of these meta-variables is associated with
 an identity substitution which can be omitted. 

We define \lstinline{Log [$\gamma$ |- M $\approx$ N] [A]} in Beluga 
as a \emph{stratified}
type, which is a form of recursive type which is defined by structural recursion on one of its indices, as an
alternative to an inductive (strictly positive) definition. Beluga verifies that this
stratification condition is satisfied. In this case, the definition is
structurally recursive on the type \lstinline{A}.

\begin{lstlisting}
stratified Log : ($\gamma$:actx) [$\gamma$ |- tmpair] -> [tp] -> ctype =
| LogBase :  [$\gamma$ |- algeqTm M N i] -> Log [$\gamma$ |- M $\approx$ N] [i]
| LogArr  : ({$\delta$:actx}{$\pi$:[$\delta$ |- $\gamma$]}{N1:[$\delta$ |- tm]}{N2:[$\delta$ |- tm]}
                Log [$\delta$ |- N1 $\approx$ N2] [T] -> Log [$\delta$ |- app M1[$\pi$] N1 $\approx$ app M2[$\pi$] N2] [S])
             -> Log [$\gamma$ |- M1 $\approx$ M2] [T $\Rightarrow$ S];
\end{lstlisting}

At base type, two terms are logically equivalent if they are algorithmically
equivalent. At arrow type we employ the monotonicity condition
mentioned in Sec. \ref{sec:proofoverview}:
\lstinline{M1} is related to \lstinline{M2} in $\Gamma$ if, for any
context $\Delta$, path substitution $\Delta \vdash \pi :
\Gamma$, and \lstinline{N1} , \lstinline{N2} related in $\Delta$, we
have that \lstinline{app M1[$\pi$] N1} is related to
\lstinline{app M2[$\pi$] N2} in $\Delta$. We quantify over
\lstinline{($\gamma$:actx)} in round parentheses, which indicates that it
is implicit and recovered during reconstruction. Variables quantified
in curly braces such as \lstinline!{$\delta$:actx}! are passed
explicitly. As in LF specifications, all free variables occurring in constructor definitions are reconstructed
and bound implicitly at the outside. They are passed implicitly and recovered
during reconstruction.

Crucially, logical equality is monotonic under path substitutions.

\begin{lstlisting}
rec log_monotone : {$\delta$:actx}{$\pi$:[$\delta$ |- $\gamma$]} Log [$\gamma$ |- M1 $\approx$ M2] [A] -> Log [$\delta$ |- M1[$\pi$] $\approx$ M2[$\pi$]] [A]
\end{lstlisting}

We show below the mechanized proof of this lemma only to illustrate the
general structure of Beluga proofs. The proof is simply by case
analysis on the logical equivalence. In the base case, we obtain a proof $P$
of \lstinline{$\gamma$ |- algeqTm M N i}, which we can weaken for free
by simply applying $\pi$ to $P$. Here we benefit significantly from
Beluga's built-in support for simultaneous substitutions; we gain not
just weakening by a single variable for free as we would in Twelf, but arbitrary simultaneous
weakening. The proof proceeds in the arrow case by simply
composing the two substitutions. We use \lstinline{mlam} as the introduction form for universal
quantifications over metavariables (contextual objects), for which we
use uppercase and Greek letters, and \lstinline{fn} with lowercase letters for
computation-level function types (implications).

\begin{lstlisting}
rec log_monotone : {$\delta$:actx}{$\pi$:[$\delta$ |- $\gamma$]} Log [$\gamma$ |- M1 $\approx$ M2] [A] -> Log [$\delta$ |- M1[$\pi$] $\approx$ M2[$\pi$]] [A] =
mlam $\delta$,$\pi$ => fn e => case e of
| LogBase [$\gamma$ |- P] => LogBase [$\delta$ |- P[$\pi$]]
| LogArr f => LogArr (mlam $\delta$',$\pi$' => f [$\delta$'] [$\delta$' |- $\pi$[$\pi$']])
\end{lstlisting}

The main lemma is mutually recursive, expressing that path
equivalence is included in logical equivalence, and logical
equivalence is included in algorithmic term equivalence. This enables
``escaping'' from the logical relation to obtain an algorithmic
equality in the end. They are
structurally recursive on the type. Crucially, in the arrow case,
\lstinline{reify} instantiates the path substitution \lstinline{$\pi$}
with a weakening substitution in order to create a fresh variable.

\begin{lstlisting}
rec reflect : {A:[tp]} [$\gamma$ |- algeqP M1 M2 A] $~\,$-> $\,$Log [$\gamma$ |- M1 $\approx$ M2] [A]
and reify   : {A:[tp]} Log [$\gamma$ |- M1 $\approx$ M2] [A] -> [$\gamma$ |- algeqTm M1 M2 A]
\end{lstlisting}

We can state weak head closure directly as follows. The proof is structurally
recursive on the type, which is implicit.

\begin{lstlisting}
rec closed : [$\gamma$ |- mstep N1 M1] -> [$\gamma$ |- mstep N2 M2] -> Log [$\gamma$ |- M1 $\approx$ M2] [T]
           -> Log [$\gamma$ |- N1 $\approx$ N2] [T]
\end{lstlisting}

\subsection{Encoding declarative equivalence}
We now define declarative equality of terms, which includes
non-algorithmic rules such as transitivity and symmetry. Declarative
equality makes use of a schema which lists only term variables, which we write
\lstinline{ctx}.

\begin{lstlisting}
schema ctx = tm;
\end{lstlisting}

For technical reasons which we will go into more detail on later, we resort to a
different treatment of typing contexts. We explicitly represent typing
contexts \lstinline{dctx} as a list of types, and declarative equality
as a computation-level inductive datatype, instead of an LF specification.

\begin{lstlisting}
LF dctx : type = 
| nil : dctx
| &   : dctx -> tp -> dctx % infix ;
\end{lstlisting}

We describe next the result of looking up the type of a variable \lstinline{x} in
$\gamma$ in typing context $\Gamma$ by its position. If \lstinline{x}
is the top variable of $\gamma$, then its type in $\Gamma$ is the type
of the top variable of $\Gamma$. Otherwise, if looking up the type of
\lstinline{x} in $\gamma$ yields \lstinline{T}, then looking it up in
an extended context also yields \lstinline{T}. Here we write
\lstinline{[$\gamma$ |- tm]} for the contextual type of terms of
type \lstinline{tm} in context $\gamma$, and \lstinline{[tp]} for
(closed) types. We use \lstinline{#p} for a meta-variable standing
for an object-level variable from $\gamma$ (as opposed to a general term).

\begin{lstlisting}
inductive Lookup : {$\Gamma$:[dctx]}($\gamma$:ctx)[$\gamma$ |- tm] -> [tp] -> ctype =
| Top : Lookup [$\Gamma$ & T] [$\gamma$,x:tm |- x] [T]
| Pop : Lookup [$\Gamma$] [$\gamma$ |- #p] [T] -> Lookup [$\Gamma$ & S] [$\gamma$,x:tm |- #p] [T];
\end{lstlisting}

We write 
\lstinline{Decl [$\Gamma$] [$\gamma$ |- M $\approx$ N] [T]}
for declarative equivalence of \lstinline{M} and \lstinline{N} at type
\lstinline{T}. We employ
the convention that $\Gamma$ and $\Delta$ stand
for typing contexts (of type \lstinline{[dctx]}), while $\gamma$ and $\delta$ stand for
corresponding term contexts.

\begin{lstlisting}
inductive Decl : {$\Gamma$:[dctx]}($\gamma$:ctx) [$\gamma$ |- tmpair] -> [tp] -> ctype =

| DecBeta : Decl [$\Gamma$ & T] [$\gamma$,x:tm |- M2 $\approx$ N2] [S] -> Decl [$\Gamma$] [$\gamma$ |- M1 $\approx$ N1] [T]
        -> Decl [$\Gamma$] [$\gamma$ |- app (lam (\x. M2)) M1  $\approx$  N2[.., N1]] [S]

| DecLam : Decl [$\Gamma$ & T] [$\gamma$,x:tm |- M $\approx$ N] [S]
        -> Decl [$\Gamma$] [$\gamma$ |- lam (\x. M)  $\approx$  lam (\x. N)] [T $\Rightarrow$ S]

| DecExt : Decl [$\Gamma$ & T] [$\gamma$,x:tm |- app M x  $\approx$  app N x] [S]
        -> Decl [$\Gamma$] [$\gamma$ |- M $\approx$ N] [T $\Rightarrow$ S]

| DecVar : Lookup [$\Gamma$] [$\gamma$ |- #p] [T] -> Decl [$\Gamma$] [$\gamma$ |- #p $\approx$ #p] [T]

| DecApp : Decl [$\Gamma$] [$\gamma$ |- M1 $\approx$ M2] [T $\Rightarrow$ S] -> Decl [$\Gamma$] [$\gamma$ |- N1 $\approx$ N2] [T]
        -> Decl [$\Gamma$] [$\gamma$ |- app M1 N1  $\approx$  app M2 N2] [S]

| DecSym : Decl [$\Gamma$] [$\gamma$ |- M $\approx$ N] [T] -> Decl [$\Gamma$] [$\gamma$ |- N $\approx$ M] [T]

| DecTrans : Decl [$\Gamma$] [$\gamma$ |- M $\approx$ N] [T] -> Decl [$\Gamma$] [$\gamma$ |- N $\approx$ O] [T]
          -> Decl [$\Gamma$] [$\gamma$ |- M $\approx$ O] [T];
\end{lstlisting}

Declarative equality includes a $\beta$ rule, as well as an extensionality rule, which states that for two terms
\lstinline{M} and \lstinline{N} to be equal at type \lstinline{T $\Rightarrow$ S}, it suffices
for them to be equal when applied to a fresh variable of type
\lstinline{T}. We again remind the reader that all meta-variables are
silently associated with the identity substitution; in particular in
\lstinline![$\gamma$ |- lam (\x.M) $\approx$ lam (\x.N)]!, the
meta-variables \lstinline!M! and \lstinline!N! are associated with the
identity substitution on the context \lstinline!$\gamma$, x:tm!. Note
that every meta-variable is associated with a simultaneous
substitutions in Beluga. If this substitution is the identity, then it can be omitted. Hence, 
\lstinline![$\gamma$ |- lam (\x.M) $\approx$ lam (\x.N)]! is
equivalent to writing 
\lstinline![$\gamma$ |- lam (\x.M[.., x]) $\approx$ lam (\x.N[..,x])]!. Written
in $\eta$-contracted form this is equivalent to:
\lstinline![$\gamma$ |- lam M $\approx$ lam N]! or making the identity
substitution explicit \lstinline![$\gamma$ |- lam M[..] $\approx$ lam N[..]]!.

Note that meta-variables associated with simultaneous substitutions do
not exist other systems. For example in LF and its implementation
in Twelf \cite{Pfenning99cade} the context of assumptions is ambient
and we cannot express dependencies of LF-variables on them. In Twelf,
writing \lstinline!lam M! is equivalent to its $\eta$-expanded form
\lstinline!lam \x. M x!.

\subsection{Fundamental theorem}

The fundamental theorem requires us to speak of all instantiations of
open terms by related substitutions. We express here the notion of
related substitutions using inductive types. Trivially, empty
substitutions, written as $\cdot$, are related at
empty domain. If $\sigma_1$ and $\sigma_2$ are related at
$\Gamma$ and \lstinline{M1} and \lstinline{M2} are related at
\lstinline{T}, then \lstinline{$\sigma_1$,M1} and
\lstinline{$\sigma_2$,M2} are related at \lstinline{$\Gamma$ & T}. The
technical reason we use the schema \lstinline{ctx} of term assumptions
only is that we would like the substitutions \lstinline{$\sigma_1$}
and \lstinline{$\sigma_2$} to carry only terms \lstinline{M}, but \emph{not}
derivations \lstinline{algeqP M M T} (or declarative equality
assumptions). If we had used the schema \lstinline{actx} or a schema
with declarative equality assumptions, the proof of the fundamental
theorem would be obligated to construct these derivations, which would
be more cumbersome.

\begin{lstlisting}
inductive LogSub : ($\gamma$:ctx)($\delta$:actx){$\sigma_1$:[$\delta$ |- $\gamma$]}{$\sigma_2$:[$\delta$ |- $\gamma$]}{$\Gamma$:[dctx]} ctype =
| Nil : LogSub [$\delta$ |- $\cdot$] [$\delta$ |- $\cdot$] [nil]
| Dot : LogSub [h |- $\sigma_1$] [h |- $\sigma_2$] [$\Gamma$] -> Log [$\delta$ |- M1 $\approx$ M2] [T]
      -> LogSub [$\delta$ |- $\sigma_1$, M1] [$\delta$ |- $\sigma_2$, M2] [$\Gamma$ & T] 
\end{lstlisting}

We have a monotonicity lemma for logically equivalent substitutions
which is similar to the monotonicity lemma for logically equivalent terms:

\begin{lstlisting}
rec wknLogSub : {$\pi$:[$\delta$' |- $\delta$]} LogSub [$\delta$ |- $\sigma_1$] [$\delta$ |- $\sigma_2$] [$\Gamma$]
                           -> LogSub [$\delta$' |- $\sigma_1$[$\pi$]] [$\delta$' |- $\sigma_2$[$\pi$]] [$\Gamma$]
\end{lstlisting}

The fundamental theorem requires a proof that \lstinline{M1} and
\lstinline{M2} are declaratively equal, together with logically
related substitutions $\sigma_1$ and $\sigma_2$, and produces a proof that
\lstinline{M1[$\sigma_1$]} and \lstinline{M2[$\sigma_2$]} are
logically related.  In the
transitivity and symmetry cases, we appeal to transitivity and symmetry
of logical equivalence, the proofs of which can be found in the
accompanying Beluga code.

\begin{lstlisting}
rec thm :  Decl [$\Gamma$] [$\gamma$ |- M1 $\approx$ M2] [T]
        -> LogSub [$\delta$ |- $\sigma_1$] [$\delta$ |- $\sigma_2$] [$\Gamma$]
        -> Log [$\delta$ |- M1[$\sigma_1$] $\approx$ M2[$\sigma_2$]] [T] =
\end{lstlisting}

We show the \lstinline{lam} case of the proof term only to make a high-level
comparison to the hand-written proof in
Sec. \ref{sec:proofoverview}. Below, one can see that we appeal to
monotonicity (\lstinline{wknLogSub}), weak head closure
(\lstinline{closed}), and the induction hypothesis on the subderivation
\lstinline{d1}. However, remarkably, there is no explicit equational reasoning
about substitutions, since applications of substitutions are
automatically simplified. We refer the reader to \cite{Cave:LFMTP13}
for the technical details of this simplification.

\begin{lstlisting}
fn d, s => case d of
| DecLam d1 =>
   LogArr (mlam $\delta'$, $\pi$, N1, N2 => fn rn =>
     let ih = thm d1 (Dot (wknLogSub [$\delta'$] [$\delta$] [$\delta'$ |- $\pi$] s) rn) in
     closed [$\delta'$ |- trans1 beta refl] [$\delta'$ |- trans1 beta refl] ih
   )
...
\end{lstlisting}

Completeness is a corollary of the fundamental theorem. Our statement
of the completeness theorem is slightly complicated by the fact that
declarative equality and algorithmic equality live in different
context schemas. To overcome this, we describe a predicate \lstinline{EmbedSub [$\Gamma$] [$\gamma$] [$\gamma$' $\vdash$ $\iota$]} which states that
$\iota$ is a simple weakening substitution which performs the work of
moving from term context \lstinline{$\gamma$:ctx} to the corresponding
(larger) algorithmic equality context \lstinline{$\gamma$':actx} with added algorithmic equality
assumptions at the types listed in \lstinline{$\Gamma$:[dctx]}. Morally,
this $\iota$ substitution plays the role of the identity substitution mentioned in
Sec. \ref{sec:proofoverview}.

\begin{lstlisting}
inductive EmbedSub : {$\Gamma$:[dctx]}{$\gamma$:ctx}($\gamma$':actx){$\iota$:[$\gamma$' |- $\gamma$]} ctype =
| INil : EmbedSub [nil] [] [$\cdot$]
| ISnoc : EmbedSub [$\Gamma$] [$\gamma$] [$\gamma$' |- $\iota$]
       -> EmbedSub [$\Gamma$ & T] [$\gamma$,x:tm] [$\gamma$',b:block x:tm,ax:algeqP x x T |- $\iota$, b.1]
\end{lstlisting}

It is then straightforward to show that embedding substitutions $\iota$
are logically related to themselves using the main lemma.

\begin{lstlisting}
rec embed_log : EmbedSub [$\Gamma$] [$\gamma$] [$\gamma$' |- $\iota$] -> LogSub [$\Gamma$] [$\gamma$] [$\gamma$' |- $\iota$] [$\gamma$' |- $\iota$]
\end{lstlisting}

The completeness theorem is stated below, and follows trivially by
composing the fundamental theorem with \lstinline{embed_log} and the main lemma to escape the logical relation.

\begin{lstlisting}
rec completeness : EmbedSub [$\Gamma$] [$\gamma$] [$\gamma$' |- $\iota$] -> Decl [$\Gamma$] [$\gamma$ |- M1 $\approx$ M2] [T]
                -> [$\gamma$' |- algeqTm M1[$\iota$] M2[$\iota$] T]
\end{lstlisting}

It is unfortunate that this transportation from
\lstinline{$\gamma$} to \lstinline{$\gamma$'} is required by the current framework of contextual types, since intuitively the
algorithmic equality assumptions in \lstinline{$\gamma$'} are completely
irrelevant for the terms \lstinline{M1} and \lstinline{M2}. It's an
open problem how to improve on this.

\subsection{Remarks}
The proof passes Beluga's typechecking and totality checking. As part of the
totality checker, Beluga performs a strict positivity check for inductive types \citep{Pientka:TLCA15,Pientka:CADE15}, and a stratification check for logical relation-style definitions.

Beluga's built-in support for simultaneous substitutions is a big win
for this proof. The proof of the monotonicity lemma is very simple, since
the (simultaneous) weakening of algorithmic equality comes for free, and there is no
need for explicit reasoning about substitution equations in the
fundamental theorem or elsewhere.  We also found that the technique of
quantifying over path substitutions as opposed to quantifying over all
extensions of a context to work surprisingly well. However, it seems
to be non-obvious when this technique will work. In an earlier version
of this proof, we had resorted to explicitly enforcing that the
substitution $\pi$ contained only \emph{variables}, limiting its
capabilities to weakening, exchange, and contraction. This was done
with an inductive datatype like the following, where the contextual type
\lstinline{#[$\delta$ |- tm]} contains only \emph{variables} of type \lstinline{tm}:

\begin{lstlisting}
datatype IsRenaming : {$\gamma$:ctx}($\delta$:ctx) {$\pi$:[$\delta$ |- $\gamma$]} ctype =
| Nil : IsRenaming [] [$\delta$ |- $\cdot$]
| Cons :  {#p:#[$\delta$ |- tm]} IsRenaming [$\gamma$] [$\delta$ |- $\pi$] -> IsRenaming [$\gamma$,x:tm] [$\delta$ |- $\pi$, #p]
\end{lstlisting}

We were surprised to learn that in fact this restriction was
unnecessary, and we could instead simply directly quantify over path
substitutions, as the schema \lstinline!actx! we rely on in our proof
already effectively restricts the substitutions we can build. However, we
suspect that the technique of explicitly restricting to renaming substitutions
may still be necessary in some cases, and that it might be convenient to have a
built-in type of these renaming-only substitutions.

We remark that the completeness theorem can in
fact be executed, viewing it as an algorithm for normalizing
derivations in the declarative system
to derivations in the algorithmic system. The extension to a proof of
decidability would be a correct-by-construction functional algorithm
for the decision problem. This is a unique feature of
carrying out the proof in a type-theoretic setting like Beluga, where the proof language also serves
as a computation language.

Some aspects of this proof could still be improved. In
particular, our treatment of the different context schemas and the
relationship between them seems unsatisfactory. We had to do a bit of
work in order to move terms from \lstinline{$\gamma$:ctx} to
\lstinline{$\gamma$':actx}, and this polluted the final statement of
the completeness theorem. It can also be difficult to
know when to resort to using an explicit context and a
computation-level datatype, like we did for
declarative equality. This suggests there is room for improvement in
Beluga's treatment of contexts, and we are exploring possible approaches.

Furthermore, one might argue that having to explicitly apply the path
substitutions $\pi$ to terms like $M[\pi]$ is somewhat unsatisfactory,
so one might wish for the ability to directly perform the bounded
quantification $\forall \Delta \geq \Gamma$ and a notion of subtyping which permits
for example \lstinline{[$\Gamma$ $\vdash$ tm] $\leq$ [$\Delta$ $\vdash$ tm]}. This is also a possibility we are exploring.

Overall, we found that that the tools provided by Beluga,
especially its support for simultaneous substitutions, worked
remarkably well to express this proof and to obviate the need for
bureaucratic lemmas about substitutions and contexts, and we are
optimistic that these techniques can scale to many other varieties of logical
relations proofs.

\section{Related Work}
Mechanizing proofs by logical relations is an excellent benchmark to evaluate
the power and elegance of a given proof development. Because it requires nested
quantification and recursive definitions, encoding logical relations has been
particularly challening for systems supporting HOAS encodings.

There are two main approaches to support reasoning about HOAS encodings: 1) In
the proof-theoretic approaches, we adopt a two-level system where we implement a
specification logic (similar to LF) inside a higher-order logic supporting
(co)inductive definitions, the approach taken in Abella
\citep{Gacek:IJCAR08}, or type theory, the aproach taken in Hybrid
\citep{Momigliano:LFMTP07}.  \LONGVERSION{Hypothetical judgments of object logics are
modelled using implication in the SL and parametric judgments are handled via
(generic) universal quantification. Substituting for an assumption is then
justified by appealing to the cut-admissibility lemma of the SL.} To distinguish
in the proof theory between quantification over variables and  quantification
over terms, \cite{Gacek:LICS08} introduce a new quantifier, $\nabla$, to
describe nominal abstraction logically. To encode logical relations one
uses recursive definitions which are part of the reasoning logic \citep{Gacek:LFMTP09}. 
Induction in these systems is typically supported by reasoning about the  height
of a proof tree; this reduces reasoning to induction over natural numbers,
although much of this complexity can be hidden in Abella. 
Compared to our development in Beluga, Abella lacks support for modelling a
context of assumptions and simultanous substitutions. As a consequence, some of
the tedious basic infrastructure to reason about open and closed terms and
substitutions still needs to be built and maintained. Moreover, Abella's
inductive proofs cannot be executed and do not yield a program for normalizing
derivations. It is also not clear what is the most effective way to
perform the quantification over all \emph{extensions} of a context in Abella.

2) The type-theoretic approaches fall into two categories: we either remain
within the logical framework and encode proofs as relations as advocated in 
Twelf  \citep{Pfenning99cade} or we build a dependently typed functional
language on top of LF to support reasoning about LF specifications as done in
Beluga. The former approach lacks logical
strength; the function space in LF is ``weak'' and only represents binding
structures instead of computations. To circumvent these limitations,
\cite{Schurmann:LICS08} proposes to implement a reasoning logic within LF and
then use it to encode logical relation arguments. This approach scales to richer
calculi \citep{Rasmussen:LFMTP13} and avoids reasoning about contexts, open terms
and simultanous substitutions explicitly. However, one might argue that it not
only requires additional work to build up a reasoning logic within
LF and prove its consistency, but is also conceptually different from what one is
used to from on-paper proofs. It is also less clear whether the approach scales
easily to proving completeness of algorithmic equality due to the need to talk
about context extensions in the definition 
of logical equivalence of terms of function type.

Outside the world of HOAS, \cite{Narboux:LFMTP08} have carried out essentially the same proof in
Nominal Isabelle, and later \cite{Urban:TOCL11} tackle the extension from the simply-typed
lambda calculus to LF. Relative to their
approach, Beluga gains substitution for free, but
more importantly, equations on substitutions are silently discharged
 by Beluga's built-in support for their equational theory, so they do not
 even appear in proofs. In contrast, proving
these equations manually requires roughly a dozen intricate lemmas.

\section{Conclusion}
Our implementation of completeness of algorithmic equality takes advantage of key infrastructure provided by Beluga: it utilizes first-class simultaneous substitutions, contexts, contextual objects and the power of recursive types. This yields a direct and compact implementation of all the necessary proofs which directly correspond to their on-paper developments. Moreover, our proof yields an executable program. 
While more work on Beluga's frontend will improve and make simpler such developments, we have demonstrated that the core language is not only suitable for standard structural induction proofs such as type safety, but also proofs by logical relations. 

\bibliographystyle{eptcs}
\bibliography{logrel-doi}

\end{document}